\documentclass[12pt,a4paper]{article}
\usepackage{amsmath}
\usepackage{graphicx}
\usepackage{amsfonts}
\usepackage{amssymb}

\newcommand{\nc}{\newcommand}
\nc{\renc}{\renewcommand}

\nc{\half}{{\textstyle{1\over2}}}

\nc{\etal}{\mbox{\it et al. }}
\nc{\ie}{{\it i.e.}}
\nc{\eg}{{\it e.g.}}

\renc{\thefootnote}{\arabic{footnote}}
\nc{\capt}[1]{{\bf Figure.} {\small\sl #1}}

\nc{\eqs}[2]{\mbox{Eqs.~(\ref{#1},\,\ref{#2})}}
\nc{\eq}[1]{\mbox{Eq.~(\ref{#1})}}

\nc{\figs}[2]{\mbox{Figs.~(\ref{#1},\,\ref{#2})}}
\nc{\fig}[1]{\mbox{Fig~.(\ref{#1})}}

\nc{\mtag}[1]{\label{#1} \mbox{\marginpar{{\footnotesize #1}}}}
\renc{\baselinestretch}{1.2}
\newlength{\overeqskip}
\newlength{\undereqskip}
\nc{\be}[1]{\begin{equation} \mbox{$\label{#1}$}}
\nc{\bea}[1]{\begin{eqnarray} \mbox{$\label{#1}$}}
\nc{\Section}[2]{\section{#2}\label{#1}}
\nc{\Bibitem}[1]{\bibitem{#1}}
\nc{\Label}[1]{\label{#1}}

\nc{\eea}{\vspace{\undereqskip}\end{eqnarray}}
\nc{\ee}{\vspace{\undereqskip}\end{equation}}
\nc{\bdm}{\begin{displaymath}}
\nc{\edm}{\end{displaymath}}
\nc{\dpsty}{\displaystyle}
\nc{\bc}{\begin{center}}
\nc{\ec}{\end{center}}
\nc{\ba}{\begin{array}}
\nc{\ea}{\end{array}}
\nc{\bab}{\begin{abstract}}
\nc{\eab}{\end{abstract}}
\nc{\btab}{\begin{tabular}}
\nc{\etab}{\end{tabular}}
\nc{\bit}{\begin{itemize}}
\nc{\eit}{\end{itemize}}
\nc{\ben}{\begin{enumerate}}
\nc{\een}{\end{enumerate}}
\nc{\bfig}{\begin{figure}}
\nc{\efig}{\end{figure}}
\nc{\arreq}{&\!=\!&}
\nc{\arrmi}{&\!-\!&}
\nc{\arrpl}{&\!+\!&}
\nc{\arrap}{&\!\!\!\approx\!\!\!&}
\nc{\non}{\nonumber\\*}

\def\lsim{\; \raise0.3ex\hbox{$<$\kern-0.75em
      \raise-1.1ex\hbox{$\sim$}}\; }
\def\gsim{\; \raise0.3ex\hbox{$>$\kern-0.75em
      \raise-1.1ex\hbox{$\sim$}}\; }
\nc{\DOT}{\hspace{-0.08in}{\bf .}\hspace{0.1in}}
\nc{\Laada}{\hbox {$\sqcap$ \kern -1em $\sqcup$}}
\nc\loota{{\scriptstyle\sqcap\kern-0.55em\hbox{$\scriptstyle\sqcup$}}}
\nc\Loota{{\sqcap\kern-0.65em\hbox{$\sqcup$}}}
\nc\laada{\Loota}
\nc{\qed}{\hskip 3em \hbox{\BOX} \vskip 2ex}

\nc{\real}{{\rm I \! R}}
\nc{\Z}{{\sf Z \!\!\! Z}}
\nc{\complex}{{\rm C\!\!\! {\sf I}\,\,}}
\def\bigid{\leavevmode\hbox{\small1\kern-3.8pt\normalsize1}}
\def\id{\leavevmode\hbox{\small1\kern-3.3pt\normalsize1}}
\nc{\slask}{\!\!\!/}
\nc{\bis}{{\prime\prime}}
\nc{\pa}{\partial}
\nc{\na}{\nabla}
\nc{\ra}{\rangle}
\nc{\la}{\langle}
\nc{\goto}{\rightarrow}
\nc{\swap}{\leftrightarrow}

\nc{\EE}[1]{ \mbox{$\cdot10^{#1}$} }
\nc{\abs}[1]{\left|#1\right|}
\nc{\at}[2]{\left.#1\right|_{#2}}
\nc{\norm}[1]{\|#1\|}
\nc{\abscut}[2]{\Abs{#1}_{\scriptscriptstyle#2}}
\nc{\vek}[1]{{\rm\bf #1}}
\nc{\integral}[2]{\int\limits_{#1}^{#2}}
\nc{\inv}[1]{\frac{1}{#1}}
\nc{\dd}[2]{{{\partial #1}\over{\partial #2}}}
\nc{\ddd}[2]{{{{\partial}^2 #1}\over{\partial {#2}^2}}}
\nc{\dddd}[3]{{{{\partial}^2 #1}\over
        {\partial #2 \partial #3}}}
\nc{\dder}[2]{{{d #1}\over{d #2}}}
\nc{\ddder}[2]{{{d^2 #1}\over{d {#2}^2}}}
\nc{\dddder}[3]{{d^2 #1}\over
        {d #2 d #3}}
\nc{\dx}[1]{d\,^{#1}x}
\nc{\dy}[1]{d\,^{#1}y}
\nc{\dz}[1]{d\,^{#1}z}
\nc{\dl}[1]{\frac{d\,^{#1}l}{(2\pi)^{#1}}}
\nc{\dk}[1]{\frac{d\,^{#1}k}{(2\pi)^{#1}}}
\nc{\dq}[1]{\frac{d\,^{#1}q}{(2\pi)^{#1}}}

\nc{\cc}{\mbox{$c.c.$ }}
\nc{\hc}{\mbox{$h.c.$ }}
\nc{\cf}{cf.\ }
\nc{\erfc}{{\rm erfc}}
\nc{\Tr}{{\rm Tr\,}}
\nc{\tr}{{\rm tr\,}}
\nc{\pol}{{\rm pol}}
\nc{\sign}{{\rm sign}}
\nc{\bfT}{{\bf T }}
\def\eV{{\rm\ eV}}

\def\TeV{{\rm\ TeV}}

\nc{\cA}{{\cal A}}
\nc{\cB}{{\cal B}}
\nc{\cD}{{\cal D}}
\nc{\cE}{{\cal E}}
\nc{\cG}{{\cal G}}
\nc{\cH}{{\cal H}}
\nc{\cL}{{\cal L}}
\nc{\cO}{{\cal O}}
\nc{\cT}{{\cal T}}
\nc{\cN}{{\cal N}}
\nc{\rvac}[1]{|{\cal O}#1\rangle}
\nc{\lvac}[1]{\langle{\cal O}#1|}
\nc{\rvacb}[1]{|{\cal O}_\beta #1\rangle}
\nc{\lvacb}[1]{\langle{\cal O}_\beta #1 |}
\nc{\bb}{\bar{\beta}}
\nc{\bt}{\tilde{\beta}}
\nc{\ctH}{\tilde{\cal H}}
\nc{\chH}{\hat{\cal H}}
\nc{\1}{\aa}
\nc{\2}{\"{a}}
\nc{\3}{\"{o}}
\nc{\4}{\AA}
\nc{\5}{\"{A}}
\nc{\6}{\"{O}}
\nc{\al}{\alpha}
\nc{\g}{\gamma}
\nc{\Del}{\Delta}
\nc{\e}{\epsilon}
\nc{\eps}{\epsilon}
\nc{\lam}{\lambda}
\nc{\om}{\omega}
\nc{\Om}{\Omega}
\nc{\ve}{\varepsilon}
\nc{\mn}{{\mu\nu}}
\nc{\vp}{\varphi}

\nc{\aap}[3]{{\it  Astron.\ Astrophys.\ }{{\bf #1} {(#2)} {#3}}}
\nc{\advp}[3]{{\it  Adv.\ in\ Phys.\ }{{\bf #1} {(#2)} {#3}}}
\nc{\annp}[3]{{\it  Ann.\ Phys.\ (N.Y.)\ }{{\bf #1} {(#2)} {#3}}}
\nc{\apl}[3]{{\it  Appl. Phys. Lett. }{{\bf #1} {(#2)} {#3}}}
\nc{\apj}[3]{{\it  Ap.\ J.\ }{{\bf #1} {(#2)} {#3}}}
\nc{\apjl}[3]{{\it  Ap.\ J.\ Lett.\ }{{\bf #1} {(#2)} {#3}}}
\nc{\app}[3]{{\it Astropart.\ Phys.\ }{{\bf #1} {(#2)} {#3}}}
\nc{\cmp}[3]{{\it  Comm.\ Math.\ Phys.\ }{{ \bf #1} {(#2)} {#3}}}
\nc{\cqg}[3]{{\it  Class.\ Quant.\ Grav.\ }{{\bf #1} {(#2)} {#3}}}
\nc{\epj}[3]{{\it  Eur.\ Phys.\ J.\ }{{\bf #1} {(#2)} {#3}}}
\nc{\epl}[3]{{\it  Europhys.\ Lett.\ }{{\bf #1} {(#2)} {#3}}}
\nc{\ijmp}[3]{{\it Int.\ J.\ Mod.\ Phys.\ }{{\bf #1} {(#2)} {#3}}}
\nc{\ijtp}[3]{{\it Int.\ J.\ Theor.\ Phys.\ }{{\bf #1} {(#2)} {#3}}}
\nc{\jmp}[3]{{\it  J.\ Math.\ Phys.\ }{{ \bf #1} {(#2)} {#3}}}
\nc{\jpa}[3]{{\it  J.\ Phys.\ A\ }{{\bf #1} {(#2)} {#3}}}
\nc{\jpc}[3]{{\it  J.\ Phys.\ C\ }{{\bf #1} {(#2)} {#3}}}
\nc{\jap}[3]{{\it J.\ Appl.\ Phys.\ }{{\bf #1} {(#2)} {#3}}}
\nc{\jpsj}[3]{{\it J.\ Phys.\ Soc.\ Japan\ }{{\bf #1} {(#2)} {#3}}}
\nc{\lmp}[3]{{\it Lett.\ Math.\ Phys.\ }{{\bf #1} {(#2)} {#3}}}
\nc{\mpl}[3]{{\it  Mod.\ Phys.\ Lett.\ }{{\bf #1} {(#2)} {#3}}}
\nc{\ncim}[3]{{\it  Nuov.\ Cim.\ }{{\bf #1} {(#2)} {#3}}}
\nc{\np}[3]{{\it  Nucl.\ Phys.\ }{{\bf #1} {(#2)} {#3}}}
\nc{\pr}[3]{{\it Phys.\ Rev.\ }{{\bf #1} {(#2)} {#3}}}
\nc{\pra}[3]{{\it  Phys.\ Rev.\ A\ }{{\bf #1} {(#2)} {#3}}}
\nc{\prb}[3]{{\it  Phys.\ Rev.\ B\ }{{{\bf #1} {(#2)} {#3}}}}
\nc{\prc}[3]{{\it  Phys.\ Rev.\ C\ }{{\bf #1} {(#2)} {#3}}}
\nc{\prd}[3]{{\it  Phys.\ Rev.\ D\ }{{\bf #1} {(#2)} {#3}}}
\nc{\prl}[3]{{\it Phys.\ Rev.\ Lett.\ }{{\bf #1} {(#2)} {#3}}}
\nc{\pl}[3]{{\it  Phys.\ Lett.\ }{{\bf #1} {(#2)} {#3}}}
\nc{\prep}[3]{{\it Phys.\ Rep.\ }{{\bf #1} {(#2)} {#3}}}
\nc{\prsl}[3]{{\it Proc.\ R.\ Soc.\ London\ }{{\bf #1} {(#2)} {#3}}}
\nc{\ptp}[3]{{\it  Prog.\ Theor.\ Phys.\ }{{\bf #1} {(#2)} {#3}}}
\nc{\ptps}[3]{{\it  Prog\ Theor.\ Phys.\ suppl.\ }{{\bf #1} {(#2)} {#3}}}
\nc{\physa}[3]{{\it  Physica\ A\ }{{\bf #1} {(#2)} {#3}}}
\nc{\physb}[3]{{\it  Physica\ B\ }{{\bf #1} {(#2)} {#3}}}
\nc{\phys}[3]{{\it Physica\ }{{\bf #1} {(#2)} {#3}}}
\nc{\rmp}[3]{{\it  Rev.\ Mod.\ Phys.\ }{{\bf #1} {(#2)} {#3}}}
\nc{\rpp}[3]{{\it Rep.\ Prog.\ Phys.\ }{{\bf #1} {(#2)} {#3}}}
\nc{\sjnp}[3]{{\it Sov.\ J.\ Nucl.\ Phys.\ }{{\bf #1} {(#2)} {#3}}}
\nc{\spjetp}[3]{{\it Sov.\ Phys.\ JETP\ }{{\bf #1} {(#2)} {#3}}}
\nc{\yf}[3]{{\it Yad.\ Fiz.\ }{{\bf #1} {(#2)} {#3}}}
\nc{\zetp}[3]{{\it Zh.\ Eksp.\ Teor.\ Fiz.\  }{{\bf #1}  {(#2)} {#3}}}
\nc{\zp}[3]{{\it Z.\ Phys.\ }{{\bf #1} {(#2)} {#3}}}
\nc{\ibid}[3]{{\sl ibid.\ }{{\bf #1} {(#2)} {#3}}}
\nc{\rf}[1]{(\ref{#1})}
\nc{\nn}{\nonumber \\*}
\nc{\bfB}{\bf{B}}
\nc{\bfv}{\bf{v}}
\nc{\bfx}{\bf{x}}
\nc{\bfy}{\bf{y}}
\nc{\vx}{\vec{x}}
\nc{\vy}{\vec{y}}
\nc{\oB}{\overline{B}}
\nc{\oI}{\overline{I}}
\nc{\oR}{\overline{R}}
\nc{\rar}{\rightarrow}
\nc{\ti}{\times}
\nc{\slsh}{\hskip-5pt/}
\nc{\sm}{Standard~Model~}
\nc{\MP}{M_{\rm Pl}}
\nc{\tp}{t_{\rm Pl}}
\nc{\ave}{\bar{E}}

\nc{\eff}{{\rm eff}}
\nc{\kk}{\vek{k}}
\nc{\pp}{{\rm p}}
\nc{\ga}{g_{a\gamma}}
\nc{\vv}{\\}
\nc{\eee}{{\bf E}}
\nc{\bbb}{{\bf B}}
\nc{\qcd}{T_{\rm QCD}}
\nc{\G}{\rm \ G}
\nc{\nom}{\nonumber}
\def\vec#1{{\bf #1}}
\begin{document}
\title{ \vskip-2truecm{\hfill {\small CFNUL/99-10}} \\
 \vskip-3mm
        {\normalsize \hfill hep-ph/9912240  } \\
\vskip 15mm
{\bf Neutrino mixing scenarios and AGN} }

\author{
Lu{\'{\i}}s Bento$^{1}$, 
Petteri Ker\"anen$^{2}$ and Jukka Maalampi$^{3}$ 
\\
{\sl\small $^{1,2}$ Centro de F{\'{\i}}sica Nuclear,
Universidade de Lisboa }\\ 
{\sl\small 
Av. Prof. Gama Pinto 2, 1649-003 Lisboa,
Portugal }\\
{\sl\small
$^{3}$ Department of Physics, P.O. Box 9,
FIN-00014 University of Helsinki,
Finland}  }

\date{December 4, 1999 \\
 {\small To appear in Phys. Lett. B} }

\maketitle
\begin{abstract}
\noindent
Active galactic nuclei (AGN) have been suggested to be sources of very
high energy neutrinos.
We consider the possibility of using AGN neutrinos to test neutrino mixings. 
From the atmospheric, solar and laboratory data on neutrino oscillations we
derive the flavour composition of the AGN neutrino flux in different neutrino 
mixing schemes. We show that most of the schemes considered
can be distinguished from each other and the existence of a sterile 
neutrino can be specially tested. AGN neutrinos can also be used to test 
those four-neutrino scenarios where solar neutrinos oscillate into an 
arbitrary mixture of $\nu_s$ and $\nu_\tau$.
\\
\\
\noindent
{\it PACS:} 14.60.Lm; 14.60.Pq; 14.60.St; 98.54.Cm

\noindent
{\it Keywords:} active galactic nuclei; neutrino mixing
\end{abstract}
\vfill %
\footnoterule
{\noindent \small $^{1}$lbento@fc.ul.pt\vskip-1pt}
{\noindent \small $^{2}$keranen@alf1.cii.fc.ul.pt\vskip-1pt}
{\noindent \small $^{3}$jukka.maalampi@helsinki.fi }

\thispagestyle{empty}
\newpage
\setcounter{page}{1}

\noindent{\it Introduction.} 
Atmospheric and solar neutrino measurements
have provided a strong evidence for the existence of neutrino oscillations, or
neutrino masses and mixings. The recent atmospheric neutrino results from
Super-Kamiokande~\cite{SKres} suggest the oscillation of the muon neutrino
($\nu_\mu$) into a tau neutrino ($\nu_\tau$) or a sterile neutrino ($\nu_s$). On the
other hand, the observed
solar neutrino deficit can be interpreted as evidence for the oscillation
of the electron neutrinos ($\nu_e$) into neutrinos of a different 
flavour (see e.g.~\cite{HataLangacker,BahcallKrastevSmirnov1}). 
In addition, the LSND collaboration has reported results of their laboratory
measurements that indicate the existence of 
$\bar\nu_\mu\rightarrow\bar\nu_e$ and 
$\nu_\mu\rightarrow\nu_e$ oscillations~\cite{LSNDoriginal}.
Unfortunately, the data does not yet uniquely fix the mass and mixing pattern among
neutrinos. A number of different scenarios are still allowed. 

The explanation
of the solar, atmospheric and LSND results requires a four-neutrino scheme of
three active neutrinos $\nu_e$, $\nu_\mu$, $\nu_\tau$ and one sterile neutrino
$\nu_s$~\cite{Juha}. The mass pattern should have a two-doublet structure~\cite{Bilenky}: 
one of the doublets
consists of $\nu_e$ and $\nu_s$ (or $\nu_e$ and $\nu_\tau$), and is responsible 
for the solar 
neutrino deficit, and the other one, responsible for the atmospheric neutrino anomaly,
consists of $\nu_\mu$ and $\nu_\tau$ (or 
$\nu_\mu$ and $\nu_s$). These
doublets are separated by a wide mass gap (${\cal O}(1 \eV)$), so as to explain the
LSND result in terms of $\nu_e - \nu_\mu$ mixing. 

There is an ambiguity in this scheme due to the fact that there are at present four 
viable interpretations of the solar neutrino data corresponding to
different choices of the oscillation parameters: 
the vacuum oscillation solution (VO),
the low MSW solution (LOW), the small mixing angle MSW solution (SMA) and 
the large mixing angle MSW solution (LMA)~\cite{HataLangacker,BahcallKrastevSmirnov1}. 
Disregarding the LSND data, it is, of course, possible to give an explanation for the other
results in terms of the known three neutrino species. 
In this case one assumes that 
the atmospheric neutrino anomaly is due to $\nu_\mu - \nu_\tau$ mixing and solar
neutrinos oscillate to $\nu_\mu$ and  $\nu_\tau$.
On the other hand, 
there might exist two (or more) sterile neutrinos ($\nu_s$, $\nu_{s'}$) so that the solar
neutrino deficit is due to $\nu_e - \nu_s$ mixing and the atmospheric 
neutrino anomaly due to $\nu_\mu - \nu_{s'}$ mixing.

In this paper we will study the possibility of testing the various neutrino 
mixing scenarios by observing the flavour composition of the high-energy
neutrino flux from active galactic nuclei (AGN). With the oscillation parameters
suggested by the solar neutrino, the atmospheric neutrino and the laboratory 
measurements, the fluxes of different neutrino flavours are determined
entirely by the mixing angles with no dependence on the
masses. It turns out that there are quite large variations of the fluxes between the different
mixing schemes, which should be detectable in the new neutrino 
telescopes~\cite{telescopes} such as AMANDA, NESTOR, BAIKAL, 
ANTARES and NEMO. With sufficient statistics, which we believe to be achievable in 
these telescopes within a reasonable time scale, one would be able to discriminate between 
different solar neutrino solutions, check the possible existence of large active-sterile 
mixings, and make distinction between three-neutrino and most of the four-neutrino schemes.


\medskip

\noindent{\it Production of AGN neutrinos.}
The AGN neutrino
production is suggested to take place in the AGN cores~\cite{coremodels},
in the jets~\cite{jetmodels} and at the endpoints of jets so-called hot
spots~\cite{hotspot}. In these sources, charged particles are supposed to
be accelerated in the vicinity of shock waves by the first order Fermi
acceleration mechanism (see e.g.~\cite{longair}). 
The collisions of the accelerated protons with photons
in the ambient electromagnetic fields would then lead to the production of 
energetic gamma rays and neutrinos via the pion photoproduction
processes~\cite{HalzenTASI},
\bea{pionphotoproduction}
p\gamma &\rightarrow & n\pi^+ \rightarrow 
\mu^+\nu_{\mu} \rightarrow e^+\nu_e\bar{\nu}_{\mu},\nom\\
p\gamma & \rightarrow & p\pi^0 \rightarrow 
2\gamma. 
\eea
The neutrino and high energy photon spectra are thus related 
to each other, and the neutrino flux can be estimated
from the high energy photon flux~\cite{HalzenTASI,Gandhi}. 
Tau neutrinos are produced in negligible amounts~\cite{PakvasaLearned}. 
Hence the flavour composition of the AGN neutrino flux is expected to be,
with a high precision,
\begin{equation}\label{initialratio}
N_\tau:N_\mu:N_e=0:2:1,
\end{equation} 
where $N_\alpha$, $\alpha=e,\mu,\tau$, denotes the total flux of $\nu_\alpha$'s and 
$\bar{\nu}_\alpha$'s.

The integrated diffuse intensity of high-energy ($E_\nu > 10\TeV$) neutrinos
from AGN and blazars detected in a detector with an effective area of
$1\ {\rm km}^2$ has been estimated to be a few hundreds per year~\cite{Gandhi}. 
When the plans to build several  km-scale
detectors become reality, the total annual flux observed will be, if these estimates turn out to be correct, of the order of
some hundreds to one thousand neutrinos. 
This is much larger than the estimated background flux of atmospheric 
neutrinos in the same energy range. On the other hand, the 
flavour composition of the atmospheric neutrino flux will be quite well understood due to 
the measurements at the lower part of the energy spectrum. Hence the background
of the atmospheric neutrinos would not crucially interfere with the determination
of the AGN neutrino fluxes. Therefore the detection of AGN neutrinos will probe new
physics beyond the Standard Model (SM), in particular neutrino oscillations~\cite{PakvasaLearned}. 

It is important to notice that 
all three neutrino flavours will be observed independently from each 
other~\cite{HalzenTASI,PakvasaLearned,Halzen}. Therefore 
any new physics beyond the SM that may alter the flavour composition
of the neutrino flux will be 
probed. Neutrino oscillations can be tested by using the flavour 
composition~\cite{PakvasaLearned}, but AGN neutrino flux is 
sensitive also to other phenomena, such as new interactions, 
magnetic moments and decays of neutrinos, which may affect the intensity
or the composition of the flux~\cite{LOT}.
Here we shall  study how the AGN neutrino data might be used to discriminate between
various neutrino mixing schemes proposed for the interpretation of the
results of the solar, atmospheric and LSND neutrino experiments.

\medskip

\noindent{\it Oscillation probabilities.}
In the four-neutrino scenarios mentioned above, the neutrinos form two pairs 
of flavours with a 
large or potentially large mixing within each pair. 
In the two basic scenarios that can be 
considered~\cite{Bilenky} all the other mixing angles are small and they
cannot substantially affect the flavour composition of the AGN neutrino flux. 
In that case it is
sufficient to consider two-flavour oscillations with either atmospheric or 
solar neutrino mixing angles
$\theta_{\rm atm}$ and $\theta_\odot$, respectively. 

The oscillation probability for the two-neutrino oscillation $\nu_\alpha\rightarrow\nu_\beta$ 
($\alpha\not=\beta$) is given by
\begin{equation}
P(\nu_\alpha \to \nu_\beta) = \frac{1}{2} \sin^2 2\theta_{\alpha\beta} 
\left(1-\cos \left(\frac{\Delta m^2 L}{2E} \right)\right),
\end{equation}
where $\Delta m^2$ is the squared mass difference of the corresponding mass eigenstates 
and $\theta_{\alpha\beta}$ is the mixing angle. 
Due to the enormous
distance $L$ to the source (typically $L\simeq 500$~Mpc) the AGN neutrino oscillations can 
be relevant to probe squared mass differences of the order $10^{-16} \eV^2$ 
for the typical AGN neutrino energies of the order of PeV.
Turning this argument around, one can say that 
for any value of $\Delta m^2$ implied by the present experimental results
the oscillating term averages to zero when integrated over the neutrino energy 
spectrum.
Obviously, matter effects could only be relevant for the neutrinos that pass across the Earth.
But, in this case, for such high energies, the weak eigenstates form effective mass eigenstates
and decouple from each others.
The transition probabilities $P(\nu_\alpha\rightarrow\nu_\beta)$ and 
$P(\nu_\alpha\rightarrow\nu_\alpha)$ are thus entirely determined by the mixing
angles $\theta_{\alpha\beta}$ between the flavour states $\nu_\alpha$ and
$\nu_\beta$:
\bea{avemixing}
P(\nu_\alpha \to \nu_\beta) = \frac{1}{2} \sin^2 2\theta_{\alpha\beta},
\
P(\nu_\alpha \to \nu_\alpha)  = 1-\frac{1}{2} \sin^2 2\theta_{\alpha\beta}\;. 
\eea

\medskip

\noindent{\it Four-neutrino models.}
 Let us first consider the four-neutrino scheme
where $\nu_\mu$ mixes 
maximally or almost maximally ($0.8\lsim \sin^2 2\theta_{\mathrm{atm}}\leq 1.0$) with $\nu_\tau$ to account for the atmospheric
neutrino anomaly while the oscillations of $\nu_e$ into sterile neutrinos
$\nu_{s}$ explain the solar neutrino deficit. Let $N_{e}=1$, $N_{\mu}=2$,
$N_{\tau}=0$ be the normalized neutrino fluxes with the relative rates as
predicted by the SM, that is, if there are no neutrino mixings. 
Then, the predicted rates for
$\nu_{e}$, $\nu_{\mu}$, $\nu_{\tau}$ and $\nu_{s}$ in this four-neutrino scenario are%
\begin{align}
N_{e}  &  =1-\frac{1}{2}\sin^{2}2\theta_{\odot}\;,\nom\\
N_{\mu}  &  =2-\sin^{2}2\theta_{\mathrm{atm}}\;,\nom\\
N_{\tau}  &  =\sin^{2}2\theta_{\mathrm{atm}}\;, \nom\\
N_{s}  &  =\frac{1}{2}\sin^{2}2\theta_{\odot}\;. \label{NNNN}
\end{align}
According to recent studies only a small MSW mixing angle
$\theta_{\odot}$ can fit the solar neutrino data with a sterile neutrino
option~\cite{HataLangacker,BahcallKrastevSmirnov1}. Thus the
$\nu_e - \nu_s$ mixing will not substantially affect
the composition of the AGN neutrino flux, and one predicts the following 
observable ratios of neutrino fluxes:%
\begin{align}
\frac{2N_{e}}{N_{\mu}+N_{\tau}}  &  =1\;,\nom\\
\frac{2N_{\tau}}{N_{\mu}+N_{\tau}}  &  =\sin^{2}2\theta_{\mathrm{atm}}\;.
\end{align}
The latter ratio would provide a new independent measurement for the 
atmospheric neutrino mixing angle $\theta_{\mathrm{atm}}$.


In the other four-neutrino scenario, the atmospheric neutrino 
anomaly is explained in terms of $\nu_{\mu} - \nu_{s}$ oscillations and the solar
neutrino deficit in terms of $\nu_{e} - \nu_{\tau}$ oscillations. In the
solar case various types of solutions can fit the data, three of them with
a large mixing angle~\cite{HataLangacker,BahcallKrastevSmirnov1}: 
the vacuum oscillation solution (VO) with 
$0.7\lsim\sin^2 2\theta_\odot\lsim 1.0$, the LOW 
solution with $0.9\lsim\sin^{2}2\theta_{\odot}\lsim1.0$ and the 
large mixing angle (LMA) MSW solution with $0.55\lsim\sin^2 2\theta \lsim 0.96$. 
The latter is now preferred as a result of the Super-Kamiokande energy 
spectrum data~\cite{BahValle}. In all these three cases, $\sin
^{2}2\theta_{\odot}$\ lies above 1/2, and this is the constraint we will use
below. 
In the small mixing angle (SMA) MSW solution 
$\sin^2 2\theta_\odot$ is very small and can be neglected.
The $\nu_e$ and the $\nu_\mu$ rates are the same as
in the previous scenario, given in (\ref{NNNN}), but the $\nu_\tau$ and the $\nu_s$ rates are 
interchanged, i.e.
\begin{align}
N_{\tau}  &  =\frac{1}{2}\sin^{2}2\theta_{\odot}\;,\nom\\
N_{s}  &  =\sin^{2}2\theta_{\mathrm{atm}}\;.
\end{align}

One interesting observable flux ratio is%
\bea{solar}
\frac{2N_{\tau}}{N_{e}+N_{\tau}}=\sin^{2}2\theta_{\odot}\;,
\eea
which provides an independent measurement of the solar neutrino
mixing angle $\theta_{\odot}$. Naturally this ratio distinguishes between the large solar mixing 
angle solutions, i.e. VO, LMA and LOW, and the SMA solution: there is no $\nu_\tau$
component in the flux in the latter case.
Another useful observable is the ratio of the $\nu_\mu$ flux to the $\nu_e$ and $\nu_\tau$ 
fluxes, for which one has the bounds 
\bea{muetau}
1\leqslant\frac{N_{\mu}}{N_{e}+N_{\tau}}=2-\sin^{2}2\theta_{\mathrm{atm}%
}\lesssim 1.2\;,
\eea
where the upper limit comes from the constraint $\sin^{2}2\theta
_{\mathrm{atm}}\gtrsim0.8$. (Note that if neutrinos did not mix this ratio
had the value 2.)

The ratio (\ref{muetau}) can be used to make a distinction between 
the scheme where the solar neutrino deficit is explained with the $\nu_e - \nu_\tau$ oscillations and the
scheme where it is done with a $\nu_e - \nu_s$ mixings. In the latter case 
(see Eqs. (\ref{NNNN})) the ratio is bounded as
\begin{equation}
\frac{1}{2}\leqslant\frac{N_{\mu}}{N_{e}+N_{\tau}}\lesssim\frac{2}{3}\;.
\end{equation}
The contrast between the two four-neutrino scenarios is evident. Let us note that 
if the solar neutrino deficit is due to large $\nu_e - \nu_s$ mixing
($\sin^2 2\theta_\odot\gsim\frac{1}{2}$), the ratio (\ref{muetau}) varies between
0.7 and 0.9. This solution is, however, strongly disfavoured by the present 
data~\cite{HataLangacker,BahcallKrastevSmirnov1}.

The predictions of the different neutrino mixing scenarios can be described in
terms of the relative 'abundances' of neutrino flavours defined as%
\begin{equation}
y_{\alpha}=\frac{3N_{\alpha}}{N_{e}+N_{\mu}+N_{\tau}}\;,\qquad\qquad
\alpha=e,\mu,\tau\;,
\end{equation}
which obey the relation $y_e+y_\mu+y_\tau=3$. In the case of no mixing (SM)
one has $y_e=1$, $y_\mu=2$ and $y_\tau=0$. 
In Fig. 1 we present the allowed regions  in
the $y_{\mu}-y_{\tau}$ plane for each of the scenarios. 
The two straight solid lines correspond to SMA MSW solutions of the solar neutrino
deficit and the two quadrangular areas with dashed lines to all of the solutions with a large
mixing angle, $0.5\lsim\sin^22\theta_\odot\leq 1$ (LMA, LOW and VO). 
The straight line labeled with  $\nu_s$ corresponds to the four-neutrino scheme
where $\nu_e-\nu_s$ mixing solves the solar neutrino deficit, and the 
other line (labeled with $\nu_{\tau}$), lying on the $y_\mu$-axis, corresponds to the scenario 
where it is explained in terms of $\nu_e-\nu_\tau$ mixing. 
The lower quadrangular area labeled with $\nu_\tau$ corresponds to the scenario where 
$\nu_e-\nu_\tau$ mixing solves the solar neutrino deficit
and the area labeled with $\nu_s$ to the scheme where it is solved by
$\nu_e-\nu_s$ mixing and the atmospheric neutrino anomaly by
$\nu_\mu-\nu_\tau$ oscillations.
As mentioned above, the latter is
disfavoured by the solar neutrino data~\cite{HataLangacker,BahcallKrastevSmirnov1}. 
Along each of the solid or dashed lines either $\theta_\odot$ or
$\theta_{\rm atm}$ is
constant. As $\sin^2 2\theta_{\rm atm}$ increases from 0.8 to 1 the abundance  
$y_\mu$ always decreases, and the growth of $\sin^22\theta_\odot$ from 0.5 to 1 is 
indicated by arrows. Hence the maximal mixing point 
$\sin^22\theta_\odot=\sin^22\theta_{\rm atm}=1$
is the upper left corner in the lower "quadrangle" and the upper right corner
in the other one. 


The basic scenarios studied so far may be considered as extreme cases of a
more general four-neutrino scenario where solar electron neutrinos oscillate not to a pure sterile $\nu_{s}$ or  a
pure active $\nu_{\tau}$ neutrino flavour but rather to some linear combination $\nu_{x}\equiv\cos
\theta\;\nu_{s}-\sin\theta\;\nu_{\tau}$ of those two. The atmospheric muon neutrinos would then 
oscillate into the corresponding orthogonal
superposition $\nu_y\equiv\sin
\theta\;\nu_{s}+\cos\theta\;\nu_{\tau}$. This kind of situation is perfectly consistent with
the limits~\cite{Bilenky} derived from the reactor disappearance experiments and with the
atmospheric neutrino observations as well. 
As to solar neutrinos, so far there does not exist a data analysis for
this 
$\nu_e-\nu_s-\nu_{\tau}$ three-neutrino mixing scheme, as complete as what 
has been done for the $\nu_e-\nu_s$ or $\nu_e-\nu_{\tau}$ 
two-neutrino oscillation
schemes~\cite{HataLangacker,BahcallKrastevSmirnov1,BahValle}.
Let $\nu_{1}$, $\nu_{2}$ be the
mass eigenstates responsible for the solar neutrino deficit and $\nu_{3}$,
$\nu_{4}$ the states involved in atmospheric neutrino oscillations. The two
pairs are separated by the LSND mass gap but no other specific mass hierarchy has to 
be assumed. The reactor experiments
constrain the mixing matrix elements $U_{e3}$, $U_{e4}$, $U_{\mu1}$, $U_{\mu
2}$ to be small~\cite{Bilenky} but not $U_{\tau i}$ or $U_{si}$. If
one drops all the mixing angles that are necessarily small (unimportant for
our purpose), the mixing matrix is given as follows:%
\begin{align}
\nu_{1} &  =     \cos\theta_{\odot}\;\nu_{e}-\sin\theta_{\odot}(\cos
\theta\;\nu_{s}-\sin\theta\;\nu_{\tau})\;,\nom\\
\nu_{2} &  =    \sin\theta_{\odot}\;\nu_{e}\;+\cos\theta_{\odot}(\cos
\theta\;\nu_{s}-\sin\theta\;\nu_{\tau})\;,\nom\\
\nu_{3} &  =    \cos\theta_{\mathrm{atm}}\;\nu_{\mu}-\sin\theta
_{\mathrm{atm}}(\sin\theta\;\nu_{s}+\cos\theta\;\nu_{\tau
})\;,\nom\\
\nu_{4} &  =     \sin\theta_{\mathrm{atm}}\;\nu_{\mu}+\cos\theta_{\mathrm{atm}%
}(\sin\theta\;\nu_{s}+\cos\theta\;\nu_{\tau})\;.
\end{align}
It is clear that the solar and atmospheric neutrino phenomena are understood
in terms of $\nu_{e} - \nu_{x}$ and $\nu_{\mu} - \nu_{y}$ oscillations, respectively 
($\nu_{x}$ and $\nu_{y}$ are the states inside brackets). 

The predicted rates of AGN neutrinos are now%
\begin{align}
N_{\tau} &  =\cos^{2}\theta\,\sin^{2}2\theta_{\mathrm{atm}}+\frac{1}{2}%
\sin^{2}\theta\,\sin^{2}2\theta_{\odot}\;,\nom\\
N_{s} &  =\sin^{2}\theta\,\sin^{2}2\theta_{\mathrm{atm}}+\frac{1}{2}\cos
^{2}\theta\,\sin^{2}2\theta_{\odot}\;,
\end{align}
while the rates of $\nu_e$ and $\nu_\mu$ remain the same 
as in Eqs.~(\ref{NNNN}).
If the angle $\theta$ is set to be an arbitrary parameter the allowed regions are enlarged
covering the two more specific schemes considered before. There are two different
regions, both limited by dotted lines: one is for solar SMA solutions and the other for large
solar mixing angle solutions LMA, LOW and VO. The angle $\theta$ varies along the dotted
lines: one has $\theta=0$ for the $\nu_e-\nu_s$ solar neutrino mixing scheme and 
$\theta=\pi/2$ for the $\nu_e-\nu_\tau$ mixing scheme. If the AGN neutrino data turns out
to lie outside the areas for these two specific schemes it will indicate that solar neutrinos
do not oscillate to pure active or sterile neutrino states but rather to a mixture of them.

\medskip

\noindent{\it Three-neutrino model.}
A radically different scenario is the one with just the three standard model 
neutrinos. That is enough for explaining the atmospheric and solar
neutrino data when  the LSND results are disregarded. It is worthwhile 
to check if the AGN high-energy neutrinos can be used to confirm or 
disprove such a three-neutrino hypothesis. 
In the viable three-neutrino scenario 
the atmospheric $\nu_\mu$'s essentially oscillate to
$\nu_\tau$  but not to $\nu_e$, as indicated by 
the Super-Kamiokande~\cite{SKres} and Chooz data~\cite{CHOOZ}. 
If $\nu_{\mu}$ and $\nu_{\tau}$ \thinspace are maximally
mixed the prediction for the AGN neutrinos is that half of the $\nu_\mu$'s 
are converted into $\nu_\tau$'s making all the observable fluxes of 
the flavours $\nu_{e}$,
$\nu_{\mu}$ and $\nu_{\tau}$ equal to each other, irrespectively to the
value of the solar neutrino mixing angle $\theta_\odot$. 
The situation changes, however, if the atmospheric neutrino mixing is 
not maximal.

In the three-neutrino scheme the solar electron neutrinos oscillate into a
particular superposition of $\nu_\mu$ and $\nu_\tau$ with which $\nu_{e}$
forms a pair of almost degenerate mass eigenstates, $\nu_{1}$ and
$\nu_{2}$, separated from the mass eigenstate $\nu_{3}$ by the mass gap
$\Delta m_{\mathrm{atm}}^{2}\simeq (2-6)\times10^{-3}\eV^2$. 
The limits obtained by the Super-Kamiokande~\cite{SKres} and Chooz~\cite{CHOOZ} 
experiments imply that $\nu_{e}$ is only a small fraction of
the mass eigenstate $\nu_{3}$, that is, the element $U_{e3}$ of
the rotation matrix is small. For our purpose this mixing can be neglected 
and will be taken to be zero. Then the matrix elements $U_{\alpha i}$ can be read from
\bea{e1}
\nu_e &=&\cos\theta_{\odot}\;\nu_{1}+\sin\theta_{\odot}\;\nu_{2}\;,\nom\\
\nu_{\mu}&=&\cos\theta_{\mathrm{atm}}(-\sin\theta_{\odot}\;\nu_{1}+\cos
\theta_{\odot}\;\nu_{2})+\sin\theta_{\mathrm{atm}}\;\nu_{3}\;,\nom\\
\nu_{\tau}&=&-\sin\theta_{\mathrm{atm}}(-\sin\theta_{\odot}\;\nu_{1}+\cos
\theta_{\odot}\;\nu_{2})+\cos\theta_{\mathrm{atm}}\;\nu_{3}\; . %
\eea
The mass eigenstates $\nu_{1}$ and $\nu_{2}$ are of course separated by the
gap $\Delta m_{\odot}^{2}$.

After many oscillation lengths the number of neutrinos with a given
flavour, $N_{\alpha}$, averages to an amount that is related to the rates
of
originally emitted neutrinos, $N_{\beta}^{0}$, as follows:%
\begin{equation}
N_{\alpha}=\sum_{i,\beta}\left|  U_{\alpha i}U_{i\beta}^{\dagger}\right|
^{2}N_{\beta}^{0}\;.
\end{equation}
Using the same normalization as before, i.e. $N_{e}^{0}=1$, $N_{\mu}^{0}=2$,
$N_{\tau}^{0}=0$, one obtains%
\bea{e2}
N_e &=&1+\frac{1}{2}\sin^{2}2\theta_{\odot}\cos2\theta_{\rm{atm}}\; ,\nom\\
N_{\mu}&=&2-\sin^{2}2\theta_{\rm{atm}}-\frac{1}{2}\sin^{2} 2\theta_{\odot
}\cos2\theta_{\rm{atm}}\cos^{2}\theta_{\rm{atm}}\; ,\nom\\
N_{\tau}&=&\sin^{2}2\theta_{\rm{atm}}-\frac{1}{2}\sin^{2}2\theta_{\odot}%
\cos2\theta_{\rm{atm}}\sin^{2}\theta_{\rm{atm}}\; .%
\eea
The total number of neutrinos is conserved and equal to 3 in this normalization. If
$\theta_{\mathrm{atm}}$\ is maximal one gets equally distributed fluxes over
all flavours, as expected. If $\theta_{\mathrm{atm}}$\ is not maximal the relative rates change and,
interestingly enough, are sensitive to the sign of $\cos2\theta_{\mathrm{atm}}$,
thus differentiating between a mixing angle $\theta_{\mathrm{atm}}$ larger
than  $\pi/4$\ from a mixing angle smaller than $\pi/4$. This can be seen in
Fig.~1, where the two small triangular-like areas correspond to $\sin^{2}2\theta
_{\mathrm{atm}}$\ between $0.8$ and $1$ and $\sin^{2}2\theta_{\odot}$\ from
$1/2$ to $1$. The upper "triangle" corresponds to a negative $\cos
2\theta_{\mathrm{atm}}$\ and the lower "triangle" to a positive $\cos
2\theta_{\mathrm{atm}}$, and they meet at the point of maximal mixing, 
$\theta_{\mathrm{atm}}=\pi/4$. Hence, 
the AGN neutrinos may be used, in principle at least, to discriminate between a negative and a
positive $\cos2\theta_{\mathrm{atm}}$. The sensitivity is best when $\sin^22\theta_{\rm atm}=0.8$
and $\sin^22\theta_\odot=1$, corresponding to the most distant points of the two "triangles". 
The small $\theta_\odot$ solution
(SMA) identifies with the straight line starting at the $(1,1)$ point, the same as for 
the four-neutrino scheme with solar SMA $\nu_e-\nu_s$ 
mixing. 
Apart from that, the three-neutrino scenario is in general quite distinct
from the four-neutrino schemes.

\medskip

\noindent{\it Exotic scenarios.} One can find in the literature still some other oscillation scenarios
in addition to the 
three-neutrino and four-neutrino models considered above, which may lead to different
predictions for the flux composition. 
Models have been proposed ~\cite{Kroli} with more than one sterile neutrino
where the solar neutrino deficit is explained in terms of
$\nu_e-\nu_s$ mixing and the atmospheric neutrino
anomaly in terms of $\nu_\mu-\nu_{s'}$ mixing, where $\nu_s$ and
$\nu_{s'}$ are two separate sterile states. 
The prediction is in that case, within the SMA solar neutrino solution,
$1.5\lsim y_\mu\lsim 1.6$ and $y_\tau =0$, the same as in the
four-neutrino scheme with solar $\nu_e-\nu_\tau$ SMA solution. 
These two cases cannot be distinguished with AGN neutrino data only.
Of course, there could also exist a 
sterile neutrino very degenarate with one of the active neutrinos, so that the
squared mass difference is below
the values probed in other phenomena. Such a neutrino could
have visible effects on the AGN neutrino flux if its mixing with one of the active neutrinos is large. 

\medskip

\noindent{\it Cosmic connection.} Let us finally note that the question of existence of sterile neutrinos and their mixings with active neutrinos 
has relevance also  for cosmology. 
Depending on the value of $\Delta m^2$, a large active-sterile neutrino
mixing could bring sterile neutrinos into thermal equilibrium
thereby increasing the effective number of light neutrinos, $N_{\nu}$~\cite{enqvist}.  According to a recent 
analysis~\cite{burles} the abundancies of light elements yield the upper bound $N_{\nu}\leq 3.20$, which sets very tight constraints on 
the active-sterile mixings, in particular on the $\nu_{\mu}-\nu_s$ mixing~\cite{shi}. The four-neutrino scenario where the 
atmospheric neutrino anomaly
is explained in terms of the $\nu_{\mu}-\nu_s$ mixing seems hence unprobable. A large  $\nu_{e}-\nu_s$ mixing is 
less stringently constained by this argument as the value of $\Delta m^2$ for the solar neutrinos is smaller than for 
the atmospheric neutrinos. The cosmological bound on  $N_{\nu}$ is a disputable matter. Some observations have indicated a higher primordial
deuterium abundancy than normally assumed, and this  leads to a less restrictive limit on the effective number of light 
neutrinos~\cite{OliveSteigmannWalker}.
The conflict with the nucleosynthesis bounds may also be avoided with a large lepton asymmetry~\cite{lasymmetry}.
The AGN flux measurements would provide a new way to check the consistency of 
this cosmological reasoning.

\medskip

\noindent{\it Conclusions.} In conclusion, if neutrinos turn out to be
produced in AGN with the rates as suggested by some
authors~\cite{coremodels,jetmodels,hotspot}, then the
measurement of the flavour composition of the AGN neutrino flux
will provide a new method to discriminate between various neutrino oscillation
schemes for solar, atmospheric and laboratory neutrinos. It would also offer a new
independent measurement of the various mixing angles, such as the atmospheric and solar
neutrino mixing angle $\theta_{\rm atm}$ and $\theta_\odot$. Of particular interest is 
the possibility
of testing the existence of sterile neutrinos and of making a distinction between small and 
large angle active-sterile mixing. Moreover, one may have a hint of the more general four-neutrino 
scenarios with large $\nu_\tau-\nu_s$ mixing, which depart from the basic solar neutrino hypothesis of $\nu_e$
oscillations into pure active or sterile neutrino states. 

The AGN neutrino data will come along with the other present and future solar, terrestrial and atmospheric 
neutrino experiments which will provide a cross-checking of neutrino mass and mixing patterns. There are 
also other possibilities to study neutrino oscillations with a long flight distance by observing the neutrino 
flux from the Galactic Center~\cite{Crocker}.

\medskip

\noindent{\it Acknowledgements.} We would like to thank Augusto Barroso for useful discussions and 
for a careful reading of the manuscript. JM wishes to thank the CFNUL for hospitality during the final 
stages of the completion of this work. PK wishes to thank the Jenny and Antti Wihuri foundation for 
financial support. This work was supported by Funda\c{c}{\~a}o para a Ci{\^e}ncia e a 
Tecnologia through the grants PRAXIS XXI/BPD/20182/99 and PESO/P/PRO/1250/98 and by 
the Academy of Finland under the project no. 40677.


\pagebreak

\begin{figure}[h]
\centering
\vspace{10pt}
\includegraphics*[width=90mm]{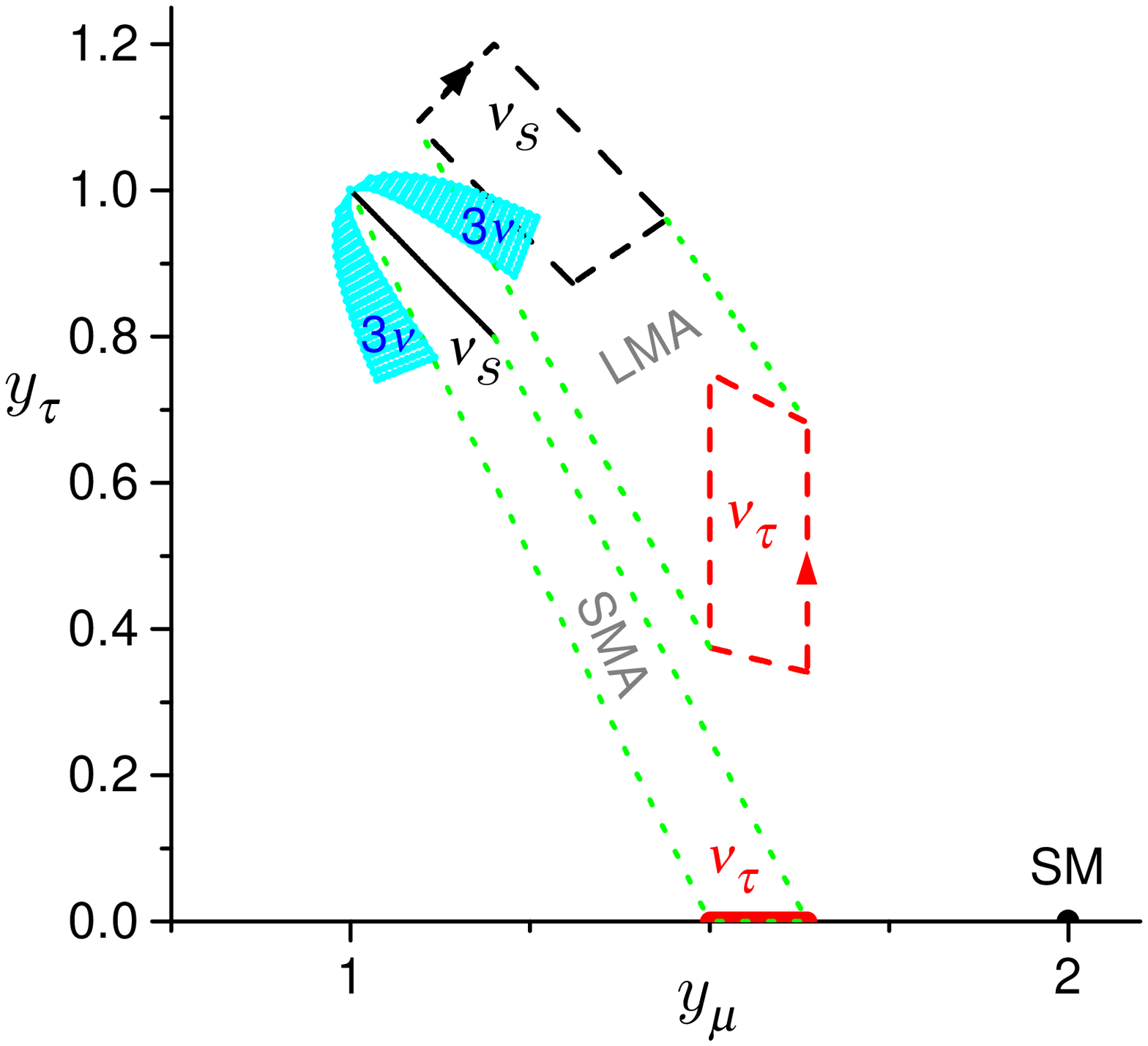}
\caption{Predictions for the rates, $y_\alpha$, of AGN $\nu_\mu$, $\nu_\tau$ and $\nu_e$
fluxes at the Earth, normalized as $y_e + y_\mu + y_\tau = 3$.
The point 'SM' corresponds to the no-mixing case.
The solid segment labeled with $\nu_s$ ($\nu_\tau$) holds for a scenario with SMA solar
$\nu_e-\nu_s$ ($\nu_e-\nu_\tau$) oscillations and atmospheric 
$\nu_\mu-\nu_\tau$ ($\nu_\mu-\nu_s$ ) oscillations.
The areas enclosed by dashed lines are associated with large 
mixing angle ($\sin^2 2\theta_\odot \ge 1/2$) solar 
neutrino solutions LMA, LOW or VO.
The case of $\nu_e-\nu_s$ oscillations 
is labeled with $\nu_s$ and the case of
$\nu_e-\nu_\tau$ oscillations is labeled with $\nu_\tau$.
The regions limited by dotted lines correspond to the four-neutrino scenario where solar
neutrinos oscillate to an arbitrary mixture of $\nu_s$ and $\nu_\tau$ which is decoupled 
from its orthogonal mixture by the LSND mass gap.
The two triangular-like areas are predicted by the LMA three-neutrino scenario, 
with $\cos2\theta_{\rm atm}<0$ in the upper "triangle" and
$\cos2\theta_{\rm atm}>0$ in the lower one. 
In all cases $\sin^2 2\theta_{\rm atm} $ varies from $0.8$ to $1$.
The arrows indicate how $\sin^2 2\theta_\odot $ grows from $1/2$ to $1$. }
\label{fig1}
\end{figure}

\end{document}